# The compositional construction of Markov processes

L. de Francesco Albasini   N. Sabadini   R.F.C. Walters

November 19, 2018

**Abstract**

We describe an algebra for composing automata in which the actions have probabilities. We illustrate by showing how to calculate the probability of reaching deadlock in $k$ steps in a model of the classical Dining Philosopher problem, and show, using the Perron-Frobenius Theorem, that this probability tends to 1 as $k$ tends to infinity.

## 1  Introduction

The idea of this paper is to introduce into the algebra of automata introduced in [4] probabilities on the actions. This permits a compositional description of probabilistic processes, in particular of Markov chains, and for this reason we call the automata we introduce *Markov automata*.

We define a Markov automaton with a given set $A$ of "signals on the left interface", and set $B$ of "signals on the right interface" to consist of a Markov matrix $\mathcal{Q}$ (whose rows and columns are the states) which is the sum

$$\mathcal{Q} = \mathsf{Q}_{a_1,b_1} + \mathsf{Q}_{a_2,b_1} + \cdots + \mathsf{Q}_{a_m,b_n}$$

of non-negative matrices $\mathsf{Q}_{a_i,b_j}$ whose elements are the probabilities of transitions between states for which the signals $a_i$ and $b_j$ occur on the interfaces. In addition, each alphabet is required to contain a special symbol $\varepsilon$ (the null signal) and the matrix $\mathsf{Q}_{\varepsilon_A,\varepsilon_B}$ is required to have row sums strictly positive.

There exist already in the literature models of probabilistic processes, for example the automata of Rabin [6], which are however non-compositional. Another model which does includes compositionality is discussed in [5]. Our model is more expressive – the example we discuss in this paper cannot be described by the model in [5]. It is also, in our view more natural, and mathematically more elegant. We will make some comparison with the cited models in the last section of the paper.

The idea of [4] was to introduce two-sided automata, in order to permit operations analogous to the parallel, series and feedback of classical circuits. For technical reasons which will become clear we first introduce weighted automata (where the weighting of a transition is a non-negative real number) and then Markov automata. We then show how to compose such automata, calculating the probabilities in composed systems. An important aspect is the use of conditional probability since, for example, composing introduces restrictions on possible transitions and hence changes probabilities.

As an illustration of the algebra we show how to specify a system of $n$ dining philosophers (a system with $12^n$ states) and to calculate the probability of reaching deadlock in $k$ steps, and we show that this probability tends to 1 as $k$ tends to $\infty$, using the methods of Perron-Frobenius theory.

It is clear that the algebra extends to semirings other than the real numbers, and in a later work we intend to discuss examples such as quantum automata. An earlier version of this paper was presented at [3].

We are grateful for helpful comments by Paweł Sobociński and Ruggero Lanotte.

## 2 Markov automata

Notice that in order to conserve symbols in the following definitions we shall use the same symbol for the automaton, its state space and its family of matrices of transitions, distinguishing the separate parts only by the font.

**Definition 2.1** *Consider two finite alphabets $A$ and $B$, containing, respectively, the symbols $\varepsilon_A$ and $\varepsilon_B$. A weighted automaton $\mathbf{Q}$ with left interface $A$ and right interface $B$ consists of a finite set $Q$ of states, and an $A \times B$ indexed family $\mathbf{Q} = (\mathsf{Q}_{a,b})_{(a \in A, b \in B)}$ of $Q \times Q$ matrices with non-negative real coefficients. We denote the elements of the matrix $\mathsf{Q}_{a,b}$ by $[\mathsf{Q}_{a,b}]_{q,q'}$ $(q, q' \in Q)$. We require further that the row sums of the matrix $\mathsf{Q}_{\varepsilon_A, \varepsilon_B}$ (and hence of $\mathcal{Q} = \sum_{a \in A, b \in B} \mathsf{Q}_{a,b}$) are strictly positive.*

We call the matrix
$$\mathcal{Q} = \sum_{a \in A, b \in B} \mathsf{Q}_{a,b}.$$
the *total matrix* of the automaton $\mathbf{Q}$.

**Definition 2.2** *Consider two finite alphabets $A$ and $B$, containing, respectively, the symbols $\varepsilon_A$ and $\varepsilon_B$. A Markov automaton $Q$ with left interface $A$ and right interface $B$, is a weighted automaton satisfying the extra condition that the row sums of the total matrix $\mathcal{Q}$ are all 1. That is, for all $q$*
$$\sum_{q'} \sum_{a \in A, b \in B} [\mathsf{Q}_{a,b}]_{q,q'} = 1.$$

*We call $[\mathsf{Q}_{a,b}]_{q,q'}$ the probability of the transition from $q$ to $q'$ with left signal $a$ and right signal $b$.*

The idea is that in a given state various transitions to other states are possible and occur with various probabilities, the sum of these probabilities being 1. The transitions that occur have effects, which we may think of a *signals*, on the two interfaces of the automaton, which signals are represented by letters in the alphabets. It is fundamental *not* to think of the letters in $A$ and $B$ as inputs or outputs, but rather signals induced by transitions of the automaton on the interfaces. For examples see section 2.3.

When both $A$ and $B$ are one element sets a Markov automaton is a Markov matrix.

**Definition 2.3** *Consider a Markov automaton* $\mathbf{Q}$ *with interfaces $A$ and $B$. A* behaviour *of length $k$ of* $\mathbf{Q}$ *consists of a two words of length $k$, one $u = a_1 a_2 \cdots a_k$ in $A^*$ and the other $v = b_1 b_2 \cdots b_k$ in $B^*$ and a sequence of non-negative row vectors*

$$x_0, x_1 = x_0 \mathsf{Q}_{a_1,b_1}, .x_2 = x_1 \mathsf{Q}_{a_2,b_2}, \cdots, x_k = x_{k-1}\mathsf{Q}_{a_k,b_k}.$$

*Notice that, in general, $x_i$ is* not *a distribution of states; for example, in our examples often $x_i = 0$.*

There is a straightforward way of converting a weighted automaton into a Markov automaton which we call *normalization*.

## 2.1 Normalization

**Definition 2.4** *The* normalization *of a weighted automaton* $\mathbf{Q}$, *denoted* $\mathbf{N}(\mathbf{Q})$ *is the Markov automaton with the same interfaces and states, but with*

$$\left[\mathsf{N}(\mathsf{Q})_{a,b}\right]_{q,q'} = \frac{[\mathsf{Q}_{a,b}]_{q,q'}}{\sum_{q' \in Q}[\mathcal{Q}]_{q,q'}} = \frac{[\mathsf{Q}_{a,b}]_{q,q'}}{\sum_{q' \in Q}\sum_{a \in A, b \in B}[\mathsf{Q}_{a,b}]_{q,q'}}.$$

To see that $\mathbf{N}(\mathbf{Q})$ is Markov, notice that the $q$th row sum of $\mathcal{N}(\mathcal{Q})$ is

$$\sum_{q'}\sum_{a,b}\left[\mathsf{N}(\mathsf{Q})_{a,b}\right]_{q,q'} = \sum_{q'}\sum_{a,b}\left[\frac{[\mathsf{Q}_{a,b}]_{q,q'}}{\sum\sum_{a,b}[\mathsf{Q}_{a,b}]_{q,q'}}\right]_{q,q'}$$
$$= \frac{\sum_{q'}\sum_{a,b}[\mathsf{Q}_{a,b}]_{q,q'}}{\sum_{q'}\sum_{a,b}[\mathsf{Q}_{a,b}]_{q,q'}} = 1.$$

**Lemma 2.5** *(i) If $\mathbf{Q}$ is a Markov automaton then $\mathbf{N}(\mathbf{Q}) = \mathbf{Q}$.*

*(ii) If $c_{a,b,q}$ are positive real numbers and $\mathbf{Q}$ and $\mathbf{R}$ are weighted automata (with the same interfaces $A$ and $B$, and the same state spaces $Q = R$) such that*

$$[\mathsf{Q}_{a,b}]_{q,q'} = c_q[\mathsf{R}_{a,b}]_{q,q'}$$

*then $\mathbf{N}(\mathbf{Q}) = \mathbf{N}(\mathbf{R})$.*

**Proof.** (ii) follows since

$$\sum_{q'}\sum_{a,b}[\mathsf{Q}_{a,b}]_{q,q'} = \sum_{q'}\sum_{a,b}c_q[\mathsf{R}_{a,b}]_{q,q'} = c_q\sum_{q'}\sum_{a,b}[\mathsf{R}_{a,b}]_{q,q'}$$

and hence

$$\frac{[\mathsf{Q}_{a,b}]_{q,q'}}{\sum_{q'}\sum_{a,b}[\mathsf{Q}_{a,b}]_{q,q'}} = \frac{c_q[\mathsf{R}_{a,b}]_{q,q'}}{c_q\sum_{q'}\sum_{a,b}[\mathsf{R}_{a,b}]_{q,q'}} = \frac{[\mathsf{R}_{a,b}]_{q,q'}}{\sum_{q'}\sum_{a,b}[\mathsf{R}_{a,b}]_{q,q'}}.$$

□

An important operation on weighted automata is the *power* construction.

## 2.2 The power construction

**Definition 2.6** *If $\mathbf{Q}$ is a weighted automaton and $k$ is a natural number, then form a weighted automaton $\mathbf{Q}^k$ as follows: the states of $\mathbf{Q}^k$ are those of $\mathbf{Q}$; the left and right interfaces are $A^k$ and $B^k$ respectively; $\varepsilon_{A^k} = (\varepsilon_A, \cdots, \varepsilon_A), \varepsilon_{B^k} = (\varepsilon_B, \cdots, \varepsilon_B)$. If $u = (a_1, a_2, \cdots, a_k) \in A^k$ and $v = (b_1, b_2, \cdots, b_k) \in B^k$ then*

$$(\mathbf{Q}^k)_{u,v} = \mathbf{Q}_{a_1,b_1} \mathbf{Q}_{a_2,b_2} \cdots \mathbf{Q}_{a_k,b_k}.$$

If $\mathbf{Q}$ is weighted and $u = (a_1, a_2, \cdots, a_k) \in A^k$, $v = (b_1, b_2, \cdots, b_k) \in B^k$, then $[(\mathbf{Q}^k)_{u,v}]_{q,q'}$ is the sum over all paths from $q$ to $q'$ with left signal sequence $u$ and right signal sequence $v$ of the weights of paths, where the weight of a path is the product of the weights of the steps.

**Lemma 2.7** *If $\mathbf{Q}$ is a weighted automaton then the total matrix of $\mathbf{Q}^k$ is the matrix power $\mathcal{Q}^k$. Hence if $\mathbf{Q}$ is Markov then so is $\mathbf{Q}^k$.*

**Proof.** The $q, q'$ entry of the total matrix of $\mathbf{Q}^k$ is

$$\sum_{u \in A^k, v \in B^k} \left[(\mathbf{Q}^k)_{u,v}\right]_{q,q'}$$

$$= \sum_{u \in A^k, v \in B^k} \left[\mathbf{Q}_{a_1,b_1} \mathbf{Q}_{a_2,b_2} \cdots \mathbf{Q}_{a_k,b_k}\right]_{q,q'}$$

$$= \sum_{u \in A^k, v \in B^k} \sum_{q_1, \cdots q_{k-1}} \left[\mathbf{Q}_{a_1,b_1}\right]_{q,q_1} \left[\mathbf{Q}_{a_2,b_2}\right]_{q_1,q_2} \cdots \left[\mathbf{Q}_{a_k,b_k}\right]_{q_{k-1},q'}$$

$$= \sum_{q_1, \cdots q_{k-1}} \sum_{a_1, \cdots, a_k} \sum_{b_1, \cdots, b_k} \left[\mathbf{Q}_{a_1,b_1}\right]_{q,q_1} \left[\mathbf{Q}_{a_2,b_2}\right]_{q_1,q_2} \cdots \left[\mathbf{Q}_{a_k,b_k}\right]_{q_{k-1},q'}$$

$$= \sum_{q_1, \cdots q_{k-1}} \sum_{a_1,b_1} \left[\mathbf{Q}_{a_1,b_1}\right]_{q,q_1} \sum_{a_2,b_2} \left[\mathbf{Q}_{a_2,b_2}\right]_{q_1,q_2} \cdots \sum_{a_k,b_k} \left[\mathbf{Q}_{a_k,b_k}\right]_{q_{k-1},q'}$$

$$= \left[\mathcal{Q}\mathcal{Q} \cdots \mathcal{Q}\right]_{q,q'}.$$

□

**Definition 2.8** *If $\mathbf{Q}$ is a Markov automaton then we call $\mathbf{Q}^k$ the automaton of $k$ step paths in $\mathbf{Q}$. We define the probability in $\mathbf{Q}$ of passing from state $q$ to $q'$ in exactly $k$ steps with left signal $u$ and right signal $v$ to be $[(\mathbf{Q}^k)_{u,v}]_{q,q'}$.*

It is important to understand the precise meaning of this definition. The probability of passing from state $q$ to $q'$ in precisely $k$ steps, so defined, is the weighted proportion of all paths of length $k$ beginning at $q$ and ending at $q'$ amongst all paths of precisely length $n$ beginning at $q$.

## 2.3 Graphical representation

Although the definitions above are mathematically straightforward, in practice a graphical notation is more intuitive. We may compress the description of an automaton with interfaces $A$ and $B$, which requires $A \times B$ matrices, into a single labelled graph, like the ones introduced in [4]. Further, expressions of automata in this algebra may be drawn as "circuit diagrams" also as in [4]. We indicate both of these matters by describing some examples.

### 2.3.1 A philosopher

Consider the alphabet $A = \{t, r, \varepsilon\}$. A philosopher is an automaton **Phil** with left interface $A$ and right interfaces $A$, state space $\{1, 2, 3, 4\}$, and transition matrices

$$\mathsf{Phil}_{\varepsilon,\varepsilon} = \begin{bmatrix} \frac{1}{2} & 0 & 0 & 0 \\ 0 & \frac{1}{2} & 0 & 0 \\ 0 & 0 & \frac{1}{2} & 0 \\ 0 & 0 & 0 & \frac{1}{2} \end{bmatrix},$$

$$\mathsf{Phil}_{t,\varepsilon} = \begin{bmatrix} 0 & \frac{1}{2} & 0 & 0 \\ 0 & 0 & 0 & 0 \\ 0 & 0 & 0 & 0 \\ 0 & 0 & 0 & 0 \end{bmatrix}, \quad \mathsf{Phil}_{\varepsilon,t} = \begin{bmatrix} 0 & 0 & 0 & 0 \\ 0 & 0 & \frac{1}{2} & 0 \\ 0 & 0 & 0 & 0 \\ 0 & 0 & 0 & 0 \end{bmatrix}$$

$$\mathsf{Phil}_{r,\varepsilon} = \begin{bmatrix} 0 & 0 & 0 & 0 \\ 0 & 0 & 0 & 0 \\ 0 & 0 & 0 & \frac{1}{2} \\ 0 & 0 & 0 & 0 \end{bmatrix}, \quad \mathsf{Phil}_{\varepsilon,r} = \begin{bmatrix} 0 & 0 & 0 & 0 \\ 0 & 0 & 0 & 0 \\ 0 & 0 & 0 & 0 \\ \frac{1}{2} & 0 & 0 & 0 \end{bmatrix}.$$

The other four transition matrices are zero matrices.

Notice that the total matrix of **Phil** is

$$\begin{bmatrix} \frac{1}{2} & \frac{1}{2} & 0 & 0 \\ 0 & \frac{1}{2} & \frac{1}{2} & 0 \\ 0 & 0 & \frac{1}{2} & \frac{1}{2} \\ \frac{1}{2} & 0 & 0 & \frac{1}{2} \end{bmatrix},$$

which is clearly stochastic, so **Phil** is a Markov automaton.

The intention behind these matrices is as follows: in all states the philosopher does a transition labelled $\varepsilon, \varepsilon$ (*idle transition*) with probability $\frac{1}{2}$; in state 1 he does a transition to state 2 with probability $\frac{1}{2}$ labelled $t, \varepsilon$ (*take the left fork*); in state 2 he does a transition to state 3 with probability $\frac{1}{2}$ labelled $\varepsilon, t$ (*take the right fork*); in state 3 he does a transition to state 4 with probability $\frac{1}{2}$ labelled $r, \varepsilon$ (*release the left fork*); and in state 4 he does a transition to state 1 with probability $\frac{1}{2}$ labelled $\varepsilon, r$ (*release the left fork*). All this information may be put in the following diagram.

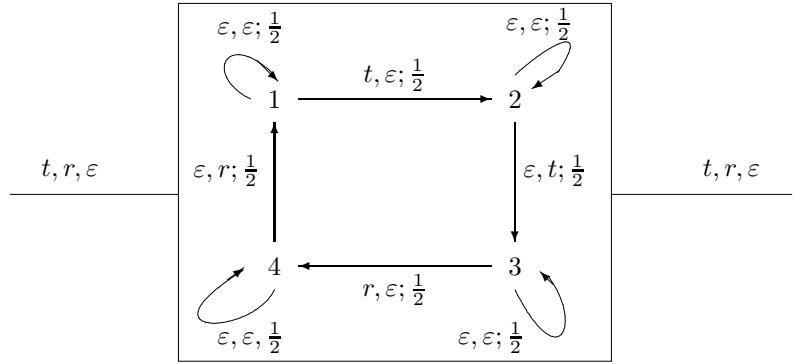

### 2.3.2 A fork

Consider again the alphabet $A = \{t, r, \varepsilon\}$. A fork is an automaton **Fork** with left interface $A$ and right interface $A$, state space $\{1, 2, 3\}$, and transition matrices

$$\mathsf{Fork}_{\varepsilon,\varepsilon} = \begin{bmatrix} \frac{1}{3} & 0 & 0 \\ 0 & \frac{1}{2} & 0 \\ 0 & 0 & \frac{1}{2} \end{bmatrix},$$

$$\mathsf{Fork}_{t,\varepsilon} = \begin{bmatrix} 0 & \frac{1}{3} & 0 \\ 0 & 0 & 0 \\ 0 & 0 & 0 \end{bmatrix}, \mathsf{Fork}_{\varepsilon,t} = \begin{bmatrix} 0 & 0 & \frac{1}{3} \\ 0 & 0 & 0 \\ 0 & 0 & 0 \end{bmatrix}$$

$$\mathsf{Fork}_{r,\varepsilon} = \begin{bmatrix} 0 & 0 & 0 \\ \frac{1}{2} & 0 & 0 \\ 0 & 0 & 0 \end{bmatrix}, \mathsf{Fork}_{\varepsilon,r} = \begin{bmatrix} 0 & 0 & 0 \\ 0 & 0 & 0 \\ \frac{1}{2} & 0 & 0 \end{bmatrix}.$$

The other four transition matrices are zero.

**Fork** is a Markov automaton since its total matrix is

$$\begin{bmatrix} \frac{1}{3} & \frac{1}{3} & \frac{1}{3} \\ \frac{1}{2} & \frac{1}{2} & 0 \\ \frac{1}{2} & 0 & \frac{1}{2} \end{bmatrix}.$$

The intention behind these matrices is as follows: in all states the fork does a transition labelled $\varepsilon, \varepsilon$ (*idle transition*) with positive probability (either $\frac{1}{3}$ or $\frac{1}{2}$); in state 1 it does a transition to state 2 with probability $\frac{1}{3}$ labelled $t, \varepsilon$ (*taken to the left*); in state 1 he does a transition to state 3 with probability $\frac{1}{3}$ labelled $\varepsilon, t$ (*taken to the right*); in state 2 he does a transition to state 1 with probability $\frac{1}{2}$ labelled $r, \varepsilon$ (*released to the left*); in state 3 he does a transition to state 1 with probability $\frac{1}{2}$ labelled $\varepsilon, r$ (*released to the right*).

All this information may be put in the following diagram:

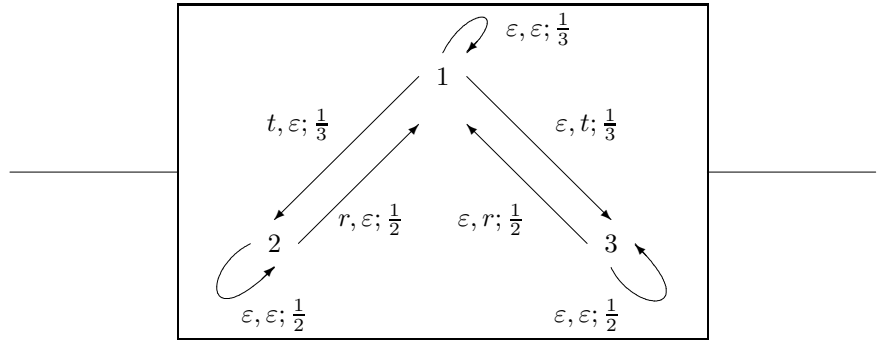

## 2.4 Reachability

For many applications we are interested only in states reachable from a given initial state by a path of positive probability. Given a Markov automaton **Q** and an initial state $q_0$ there is a subautomaton **Reach**(**Q**,$q_0$) whose states are the reachable states, and whose transitions are those of **Q** restricted to the reachable states.

# 3 The algebra of Markov automata: operations

Now we define operations on weighted automata analogous (in a precise sense) to those defined in [4].

**Definition 3.1** *Given weighted automata $\mathbf{Q}$ with left and right interfaces $A$ and $B$, and $\mathbf{S}$ with interfaces $C$ and $D$ the* parallel composite $\mathbf{Q} \times \mathbf{R}$ *is the weighted automaton which has set of states $Q \times R$, left interfaces $A \times C$, right interface $B \times D$, $\varepsilon_{A \times C} = (\varepsilon_A, \varepsilon_C)$, $\varepsilon_{B \times D} = (\varepsilon_B, \varepsilon_D)$ and transition matrices*

$$(\mathsf{Q} \times \mathsf{S})_{(a,c),(b,d)} = \mathsf{Q}_{a,b} \otimes \mathsf{S}_{c,d}.$$

This just says that the weight of a transition from $(q, r)$ to $(q', r')$ with left signal $(a, c)$ and right signal $(b, d)$ is the product of the weights of the transition $q \to q'$ with signals $a$ and $b$, and the weight of the transition $r \to r'$ with signals $c$ and $d$.

**Lemma 3.2** *If $\mathbf{Q}$ and $\mathbf{R}$ are weighted automata then*

$$\mathbf{N}(\mathbf{Q} \times \mathbf{R}) = \mathbf{N}(\mathbf{Q}) \times \mathbf{N}(\mathbf{R}).$$

*Hence if $\mathbf{Q}$ and $\mathbf{R}$ are Markov automata then so is $\mathbf{Q} \times \mathbf{R}$.*

**Proof.**

$$\begin{aligned}
\left[\mathsf{N}(\mathsf{Q} \times \mathsf{R})_{(a,c),(b,d)}\right]_{(q,r),(q',r')} &= \frac{[\mathsf{Q}_{a,b}]_{q,q'} [\mathsf{S}_{c,d}]_{r,r'}}{\sum_{q',r'} \sum_{(a,c),(b,d)} [\mathsf{Q}_{a,b}]_{q,q'} [\mathsf{S}_{c,d}]_{r,r'}} \\
&= \frac{[\mathsf{Q}_{a,b}]_{q,q'} [\mathsf{S}_{c,d}]_{r,r'}}{\sum_{q',(a,b)} [\mathsf{Q}_{a,b}]_{q,q'} \sum_{r',(c,d)} [\mathsf{S}_{c,d}]_{r,r'}} \\
&= \left[\mathsf{N}(\mathsf{Q})_{(a,b)}\right]_{q,q'} \left[\mathsf{N}(\mathsf{R})_{(c,d)}\right]_{r,r'} \\
&= \left[(\mathsf{N}(\mathsf{Q}) \times \mathsf{N}(\mathsf{R}))_{(a,c),(b,d)}\right]_{(q,r),(q',r')}.
\end{aligned}$$

For the second part notice that if $\mathbf{Q}$ and $\mathbf{R}$ are Markov then

$$\mathbf{Q} \times \mathbf{R} = \mathbf{N}(\mathbf{Q}) \times \mathbf{N}(\mathbf{R}) = \mathbf{N}(\mathbf{Q} \times \mathbf{R})$$

which implies that $\mathbf{Q} \times \mathbf{R}$ is Markov.
□

**Definition 3.3** *Given weighted automata $\mathbf{Q}$ with left and right interfaces $A$ and $B$, and $\mathbf{R}$ with interfaces $B$ and $C$ the* series (communicating parallel) *composite of weighted automata $\mathbf{Q} \circ \mathbf{R}$ has set of states $Q \times R$, left interfaces $A$, right interface $C$, and transition matrices*

$$(\mathsf{Q} \circ \mathsf{R})_{a,c} = \sum_{b \in B} \mathsf{Q}_{a,b} \otimes \mathsf{R}_{b,c}.$$

**Lemma 3.4** $(\mathbf{Q} \circ \mathbf{R}) \circ \mathbf{S} = \mathbf{Q} \circ (\mathbf{R} \circ \mathbf{S})$.

**Proof.** This follows from the fact that $\otimes$ is associative..
$\square$

It is easy to see that $\mathbf{Q}\circ\mathbf{R}$ is not necessarily Markov even when both $\mathbf{Q}$ and $\mathbf{R}$ are. The reason is that the communication in the series composite reduces the number of possible transitions, so that we must normalize to get (conditional) probabilities. However in a multiple composition it is only necessary to normalize at the end, because of the following lemma.

**Lemma 3.5** $\mathbf{N}(\mathbf{N}(\mathbf{Q})\circ\mathbf{N}(\mathbf{R})) = \mathbf{N}(\mathbf{Q}\circ\mathbf{R})$.

**Proof.**

$$\left[(\mathsf{N}\mathsf{Q}\circ\mathsf{N}\mathsf{R})_{a,c}\right]_{(q,r),(q',r')} = \sum_{b\in B} \left[\mathsf{N}\mathsf{Q}_{a,b}\right]_{q,q'} \otimes \left[\mathsf{N}\mathsf{R}_{b,c}\right]_{r,r'}$$

$$= \sum_{b\in B} \frac{[\mathsf{Q}_{a,b}]_{q,q'}}{\sum_{q'}\sum_{a,b}[\mathsf{Q}_{a,b}]_{q,q'}} \cdot \frac{[\mathsf{R}_{b,c}]_{r,r'}}{\sum_{r'}\sum_{b,c}[\mathsf{R}_{b,c}]_{r,r''}}$$

$$= \frac{1}{\sum_{q',r'}(\sum_{a,b}[\mathsf{Q}_{a,b}]_{q,q'}\sum_{b,c}[\mathsf{R}_{b,c}]_{r,r'})} \sum_{b\in B} [\mathsf{Q}_{a,b}]_{q,q'} [\mathsf{R}_{b,c}]_{r,r'}.$$

$$= c_{q,r}\left[(\mathsf{Q}\circ\mathsf{R})_{a,c}\right]_{(q,r),(q',r')}, \text{ where}$$

$$c_{q,r} = \frac{1}{\sum_{q',r'}(\sum_{a,b}[\mathsf{Q}_{a,b}]_{q,q'}\sum_{b,c}[\mathsf{R}_{b,c}]_{r,r''})} \text{ depends only on } q,r.$$

Hence by the lemma 2.5 above

$$\mathbf{N}(\mathbf{N}\mathbf{Q}\circ\mathbf{N}\mathbf{R}) = \mathbf{N}(\mathbf{Q}\circ\mathbf{R}).$$

$\square$

**Definition 3.6** *If $\mathbf{Q}$ and $\mathbf{R}$ are Markov automata, $\mathbf{Q}$ with left interface $A$ and right interface $B$, $\mathbf{R}$ with left interface $B$ and right interface $C$ then the series composite of Markov automata $\mathbf{Q}\cdot\mathbf{R}$ is defined to be $\mathbf{Q}\cdot\mathbf{R} = \mathbf{N}(\mathbf{Q}\circ\mathbf{R})$.*

**Theorem 3.7** $(\mathbf{Q}\cdot\mathbf{R})\cdot\mathbf{S} = \mathbf{Q}\cdot(\mathbf{R}\cdot\mathbf{S})$.

**Proof.**

$$(\mathbf{Q}\cdot\mathbf{R})\cdot\mathbf{S} = \mathbf{N}(\mathbf{N}(\mathbf{Q}\circ\mathbf{R})\circ\mathbf{S})$$
$$= \mathbf{N}(\mathbf{N}(\mathbf{Q}\circ\mathbf{R})\circ\mathbf{N}(\mathbf{S})) \text{ since } \mathbf{S} \text{ is Markov}$$
$$= \mathbf{N}((\mathbf{Q}\circ\mathbf{R})\circ\mathbf{S}) = \mathbf{N}(\mathbf{Q}\circ(\mathbf{R}\circ\mathbf{S})) \text{ by 3.6 and 3.6}$$
$$= \mathbf{N}(\mathbf{N}(\mathbf{Q})\circ\mathbf{N}(\mathbf{R}\circ\mathbf{S})) \text{ by 3.6}$$
$$= \mathbf{N}(\mathbf{Q}\circ\mathbf{N}(\mathbf{R}\circ\mathbf{S})) = \mathbf{Q}\cdot(\mathbf{R}\cdot\mathbf{S}) \text{ since } \mathbf{Q} \text{ is Markov.}$$

$\square$

**Theorem 3.8** *If $\mathbf{Q}$ and $\mathbf{R}$ are Markov automata then*

$$(\mathbf{Q}\times\mathbf{R})^k = \mathbf{Q}^k \times \mathbf{R}^k$$

**Proof.**

$$(Q \times R)^k)_{(u,u'),(v,v')}$$
$$= (Q_{a_1,b_1} \otimes R_{a'_1,b'_1})(Q_{a_2,b_2} \otimes R_{a'_2,b'_2}) \cdots (Q_{a_k,b_k} \otimes R_{a'_k,b'_k})$$
$$= (Q_{a_1,b_1} \otimes Q_{a_2,b_2} \otimes \cdots Q_{a_k,b_k})(R_{a'_1,b'_1} \otimes R_{a'_2,b'_2} \cdots \otimes R_{a'_k,b'_k})$$
$$= (Q^k \times R^k)_{(u,u'),(v,v')}.$$

□

**Remark.** It is not the case that if $\mathbf{Q}$, $\mathbf{R}$ are Markov then $(\mathbf{Q} \cdot \mathbf{R})^k = \mathbf{Q}^k \cdot \mathbf{R}^k$. The reason is that normalizing length $k$ steps in a weighted automaton is not the same as considering $k$ step paths in the normalization of the automaton.

Next we define some constants each of which is a Markov automaton.

**Definition 3.9** *Given a relation $\rho \subset A \times B$ such that $(\varepsilon_A, \varepsilon_B) \in \rho$ we define a Markov automaton $\rho$ as follows: it has one state $*$ say. The transition matrices $[\rho_{a,b}]$ are $1 \times 1$ matrices, that is, real numbers. Let $|\rho|$ be the number of elements in $\rho$. Then $\rho_{a,b} = \frac{1}{|\rho|}$ if $\rho$ relates $a$ and $b$, and $\rho_{a,b} = 0$ otherwise.*

Some special cases, all described in [4], have particular importance:

(i) the automaton corresponding to the identity function $1_A$, considered as a relation on $A \times A$ is called $\mathbf{1}_A$;

(ii) the automaton corresponding to the diagonal function $\Delta : A \to A \times A$ (considered as a relation) is called $\mathbf{\Delta}_A$; the automaton corresponding to the opposite relation of $\Delta$ is called $\nabla_A$.

(iii) the automaton corresponding to the function $twist : A \times B \to B \times A$ is called $\mathbf{twist}_{A,B}$.

(iv) the automaton corresponding to the relation $\eta = \{(*,(a,a); a \in A\} \subset \{*\} \times (A \times A)$ is called $\eta_A$; the automaton corresponding to the opposite of $\eta$ is called $\epsilon_A$.

### 3.1 The dining philosophers system

Now the model of the dining philosophers problem we consider is an expression in the algebra, involving also the automata **Phil** and **Fork**. The the system of $n$ dining philosophers is

$$\mathbf{DF}_n = \eta_A \cdot ((\mathbf{Phil} \cdot \mathbf{Fork} \cdot \mathbf{Phil} \cdot \mathbf{Fork} \cdot \cdots \cdot \mathbf{Phil} \cdot \mathbf{Fork}) \times \mathbf{1}_A) \cdot \epsilon_A,$$

where in this expression there are $n$ philosophers and $n$ forks.

As explained in [4], we may represent this system by the following diagram, where we abbreviate **Phil** to **P** and **Fork** to **F**.

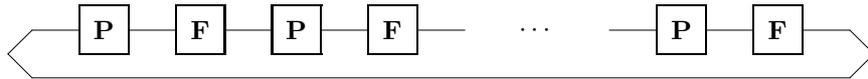

Let us examine the case when $n = 2$ with initial state $(1, 1, 1, 1)$. Let $\mathbf{Q}$ be the reachable part of $\mathbf{DF}_2$. The states reachable from the initial state are $q_1 = (1, 1, 1, 1)$, $q_2 = (1, 3, 3, 2)$, $q_3 = (3, 2, 1, 3)$, $q_4 = (1, 1, 4, 2)$, $q_5 = (4, 2, 1, 1)$, $q_6 = (1, 3, 2, 1)$, $q_7 = (2, 1, 1, 3)$, $q_8 = (2, 3, 2, 3)$ ($q_8$ is the unique deadlock state). The single matrix of the automaton $\mathsf{Q}$, using this ordering of the states, is

$$\begin{bmatrix} \frac{1}{4} & 0 & 0 & 0 & 0 & \frac{1}{4} & \frac{1}{4} & \frac{1}{4} \\ 0 & \frac{1}{2} & 0 & \frac{1}{2} & 0 & 0 & 0 & 0 \\ 0 & 0 & \frac{1}{2} & 0 & \frac{1}{2} & 0 & 0 & 0 \\ \frac{1}{2} & 0 & 0 & \frac{1}{2} & 0 & 0 & 0 & 0 \\ \frac{1}{2} & 0 & 0 & 0 & \frac{1}{2} & 0 & 0 & 0 \\ 0 & \frac{1}{3} & 0 & 0 & 0 & \frac{1}{3} & 0 & \frac{1}{3} \\ 0 & 0 & \frac{1}{3} & 0 & 0 & 0 & \frac{1}{3} & \frac{1}{3} \\ 0 & 0 & 0 & 0 & 0 & 0 & 0 & 1 \end{bmatrix}.$$

Calculating powers of this matrix we see that the probability of reaching deadlock from the initial state in 2 steps is $\frac{23}{48}$, in 3 steps is $\frac{341}{576}$, and in 4 steps is $\frac{4415}{6912}$.

## 4 The probability of deadlock

The idea of this section is to apply Perron-Frobenius theory (see, for example) [7] to the Dining Philosopher automaton. However, for convenience, we give the details of the proof of the case we need, without refering to the general theorem.

**Definition 4.1** *Consider a Markov automaton $\mathbf{Q}$ with input and output sets being one element sets $\{\varepsilon\}$. A state $q$ is called a* deadlock *if the only transition out of $q$ with positive probability is a transition from $q$ to $q$ (the probability of the transition must necessarily be 1).*

**Theorem 4.2** *Consider a Markov automaton $\mathbf{Q}$ with interfaces being one element sets, with an initial state $q_0$. Suppose that*

*(i) $\mathbf{Q}$ has precisely one reachable deadlock state,*

*(ii) for each reachable state, not a deadlock, there is a path with non-zero probability to $q_0$, and*

*(iii) for each reachable state $q$ there is a transition with non-zero probability to itself.*

*Then the probability of reaching a deadlock from the initial state in $k$ steps tends to 1 as $k$ tends to infinity.*

**Proof.** Let $\mathsf{R} = \mathbf{Reach}(\mathbf{Q}, q_0)$. Suppose $\mathsf{R}$ has $m$ states. Then in writing the matrix $\mathsf{R}$ we choose to put the deadlock last, so that $\mathsf{R}$ has the form

$$\mathsf{R} = \begin{bmatrix} \mathsf{S} & \mathsf{T} \\ 0 & 1 \end{bmatrix}$$

where $\mathsf{S}$ is $(m-1) \times (m-1)$ and $\mathsf{T}$ is $(m-1) \times 1$. Now

$$\mathsf{R}^k = \begin{bmatrix} \mathsf{S}^k & \mathsf{T}_k \\ 0 & 1 \end{bmatrix}$$

for some matrix $\mathsf{T}_k$. Condition (i) implies that there is a path with positive probability (a positive path) from any non-deadlock state to any other in **R**. Condition (ii) implies that if there is a positive path of length $l$ between two states then there is also a positive path of all lengths greater than $l$. These two facts imply that there is a $k_0$ such that from any non-deadlock state to any other state there is a positive path of length $k_0$. For this $k_0$ the matrix $T_{k_0}$ is *strictly positive*. This means that the row sums of $\mathsf{S}^{k_0}$ are strictly less than 1. But the eigenvalues of a matrix are dominated in absolute value by the maximum of the absolute row sums (the sums of absolute values of the row elements). Hence the eigenvalues of $\mathsf{S}^{k_0}$ and hence of $\mathsf{S}$ all have absolute value less than 1. But by considering the Jordan canonical form of a matrix whose eigenvalues all have absolute values less than 1 it is easy to see that $\mathsf{S}^k$ tends to 0 as $k$ tends to infinity. Hence $\mathsf{T}_k$ tends to the column vector all of whose entries are 1. Hence the probability of reaching the deadlock from any of the other states in $k$ steps tends to 1 as $k$ tends to infinity.
□

**Corollary 4.3** *In the dining philosopher problem* $\mathbf{DF}_n$ *with $q_0$ being the state* $(1, 1, \cdots, 1)$ *the probability of reaching a deadlock from the initial state in $k$ steps tends to* 1 *as $k$ tends to infinity.*

**Proof.** We just need to verify the conditions of the theorem for the dining philosopher problem. It is straightforward to check that the state

$$(2, 3, 2, 3, \cdots, 2, 3)$$

in which the philosophers are all in state 2 and the forks in state 3 is a reachable deadlock. It is clear that in any state $q$ there is a positive transition to $q$, since each component has silent moves in each state. We need only check that for any reachable state other than this deadlock that there is a positive path to the initial state. Consider the states $f_1$, $f_2$ of two forks adjacent modulo $n$, and the state $p$ of the philosopher between these two forks. Examining the positive paths possible in two adjacent forks and the corresponding philosopher we see that the reachable configurations are limited to (a) $f_1 = 1$, $p = 1$, $f_2 = 1$, (b) $f_1 = 1$, $p = 1$, $f_2 = 3$, (c) $f_1 = 1$, $p = 4$, $f_2 = 2$, (d) $f_1 = 2$, $p = 1$, $f_2 = 1$, (e) $f_1 = 2$, $p = 1$, $f_2 = 3$, (f) $f_1 = 3$, $p = 2$, $f_2 = 1$, (g) $f_1 = 3$, $p = 2$, $f_2 = 3$, (h) $f_1 = 3$, $p = 3$, $f_2 = 2$, (i) $f_1 = 2$, $p = 4$, $f_2 = 2$. We will show that in states other than the deadlock or the inital state there is a transition of the system which increases the number of forks in state 1. Notice that in a reachable state the states of adjacent forks determine the state of the philosopher between. Consider the possible configurations of fork states. We need not consider cases all forks are in state 1 (initial), or all in state 3 (the known deadlock). Given two adjacent forks in states $3, 2$ there are transitions which only involve this philosopher and the two forks (apart from null signals) which result in one of the forks returning to state 1 (the philosopher puts down a fork that he holds). This is also the case when two adjacent forks are in states $1, 2$ or $2, 2$ or $3, 1$. But in a circular arrangement other than all 1's or all 3's one of the pairs $1, 2$ or $2, 2$ or $3, 1$ or $3, 2$ must occur.
□

**Remark.** Notice that in the proof of the corollary we did not use the specific positive probabilities of the actions of the philosophers and forks. Hence the result is true with any positive probabilities replacing the specific ones we gave in the description of the philosopher and fork. In fact, different philosophers and forks may have different probabilities without affecting the conclusion of the corollary.

## 5 Concluding remarks

### 5.1 The algebra of automata: equations

There is much more to say about the algebraic structure and its relations with other fields. We have mentioned above some equations which are satisfied, and here we mention one more.

**Lemma 5.1** *The constants $\boldsymbol{\Delta}_A, \nabla_A$ satisfy the Frobenius equations [1], namely that*
$$(\boldsymbol{\Delta}_A \times \mathbf{1}_A) \cdot (\mathbf{1}_A \times \nabla_A) = \nabla_A \cdot \boldsymbol{\Delta}_A.$$

The proof is straightforward.

### 5.2 Comparisons

According to Rabin a *probabilistic automaton* on an alphabet $\Sigma$ consists of a set of states $Q$ and a family of stochastic transition matrices $[P_a]_{q,q'}$ ($a \in \Sigma$; $q, q' \in Q$). A *distribution of states* is a row vector with non negative real entries whose sum is 1. A behaviour corresponding to an initial state distribution $x_0$, and an input word $u = a_1 a_2 \cdots a_k$, is a sequence of state distributions
$$x_0, x_1 = x_0 P_{a_1}, x_2 = x_0 P_{a_2}, \cdots, x_k = x_{k-1} P_{a_k}.$$

This is a non-compositional model, and it immediately clear that the meaning of the alphabet in Rabin is quite different from the meaning of the alphabets for our Markov automata. For Rabin the letters are inputs which drive the automaton $\mathbf{Q}$ – for a given state $q$ and a given input $a$ the sum of the probabilities of transitions out of $q$ is 1. We are able to describe the same phenomenon by considering a second automaton $\mathbf{R}$ whose signal on the interface drive $\mathbf{Q}$, which of course introduces conditional probabilities. From our point of view Rabin's probabilities are conditional ones resulting from the knowledge that an input $a$ occurs. Another difference is that every transition in every state in Rabin's automata produces a distribution of states. However it is crucial in our model that actions are not necessarily defined in all states; or if they are defined they may be only partially defined. For example, the fork in state 2 has no transitions labelled $t, \varepsilon$; it cannot be taken again when it is already taken. The fork in state 1 may be taken to the left with probability $\frac{1}{2}$.

The second model we mentioned [5] considers a generalization of Rabin's model (and hence different from ours) which is influenced by concurrency theory. It has a form of composition, and as usual with models related to process algebras the composition involves an underlying broadcast (and hence interleaved) communication, and does not involve conditional probability. As a result it is not possible to describe our example, in which, in a single step all philosophers may take their left fork. Instead, it is straightforward to model broadcast, interleaved models using our algebra [2], using in particular the component $\Delta$.

Further, [5] has a much more limited algebra than that presented here; for example, multiply simultaneous signals (together with the synchronization on some of the signals) are not available.

In our view interleaving destroys the realism of the model. For example, to reach deadlock in the dining philosopher problem requires a sequence of actions, as philosophers take the forks one by one. This results in quite different probabilities.